\documentclass[twocolumn,pre,floatfix,superscriptaddress,longbibliography]{revtex4}
\usepackage{color}
\usepackage{tabularx}
\usepackage[utf8]{inputenc}
\usepackage{amsmath,amssymb}
\usepackage{makecell}
\usepackage{dblfloatfix}
\usepackage{fixltx2e}
\usepackage{epsfig}
\usepackage{amsmath}
\usepackage{soul}
\usepackage{lipsum}
\usepackage{amssymb}
\usepackage{bm}
\usepackage{soul}
\usepackage{graphicx}
\usepackage{multirow}
\usepackage{wasysym}
\usepackage{varwidth}
\usepackage{enumitem}
\usepackage{algorithm, algpseudocode}
\usepackage{epsfig}
\usepackage{lipsum}
\usepackage{amsmath}
\usepackage[caption=false, justification=centerlast]{subfig}
\usepackage[svgnames]{xcolor}
\usepackage{hyperref}
\usepackage{times}
\usepackage{etoolbox}
\hypersetup{colorlinks=true,
            linkcolor=blue,
            citecolor=ForestGreen,
            urlcolor=DarkOrchid
}

\newcommand{\fe}{\tilde{F}}
\newcommand{\de}{\mathcal{E}}
\newcommand{\bn}{\beta_n}
\newcommand{\lamb}{\boldsymbol{\lambda}}
\newcommand{\Lamb}{\boldsymbol{\Lambda}}
\newcommand{\Xt}{\tilde{X}}
\newcommand{\Xb}{\mathbf{X}}

\newcommand{\obs}{A}

\newcommand{\PA}{PA\,}
\newcommand{\SA}{SA\,}
\newcommand{\AIS}{AIS\,}
\newcommand{\x}{\sigma}
\newcommand{\Rs}{R^*}
\newcommand{\rl}{\eta}
\newcommand{\rln}{\eta(\beta_n)}
\newcommand{\prr}{PRR}
\newcommand{\wt}{w}
\DeclareMathOperator*{\E}{\mathbb{E}}
\DeclareMathOperator*{\var}{\mathrm{var}}
\DeclareMathOperator{\Cov}{Cov}
\makeatletter
\patchcmd{\math@cr@@@align}% <cmd>
  {\place@tag}% <search>
  {\bgroup\color{black}\place@tag\egroup}% <replace>
  {}{}% <success><failure>
\makeatother
\begin{document}
\title{Optimal schedules for annealing algorithms}

\author{Amin Barzegar}
\affiliation{Microsoft Quantum, Microsoft, Redmond, Washington 98052, USA}
\author{Firas Hamze}
\affiliation{Microsoft Quantum, Microsoft, Redmond, Washington 98052, USA}
\author{Christopher Amey}
\affiliation{Department of Physics, Brandeis University, Waltham, Massachusetts 02453, USA}
\author{Jonathan Machta}
\affiliation{Department of Physics, University of Massachusetts, Amherst, Massachusetts 01003, USA}
\affiliation{Santa Fe Institute, 1399 Hyde Park Road, Santa Fe, New Mexico 87501, USA}

\begin{abstract}
Annealing algorithms such as simulated annealing and population annealing are widely used both for sampling the Gibbs distribution and  solving  optimization problems (i.e. finding ground states).  For both statistical mechanics and optimization, additional parameters beyond temperature  are often needed such as chemical potentials, external fields or Lagrange multipliers enforcing constraints.  In this paper we derive a formalism for optimal annealing schedules in multidimensional parameter spaces using methods from non-equilibrium statistical mechanics.  The results are closely related to work on optimal control of thermodynamic systems [Sivak and Crooks, PRL 108, 190602 (2012)].    Within the formalism, we compare the efficiency of population annealing and multiple weighted runs of simulated annealing (“annealed importance sampling”) and discuss the effects of non-ergodicity on both algorithms.  Theoretical results are supported by numerical simulations of spin glasses.
\end{abstract}
\date{\today}
\maketitle

\section{Introduction}

Simulated annealing \cite{KiGeVe83} and related Monte Carlo algorithms such as annealed importance sampling \cite{Neal2001}, population annealing \cite{HuIb03,Mac10a}, parallel tempering \cite{Geyer91,HuNe96}, simulated tempering \cite{MaPa}, the Wang-Landau algorithm \cite{WaLa01} and nested sampling \cite{Skilling06} 
play an important role both in equilibrium sampling and optimization, and have been widely applied and studied in physics, mathematics, computer science and operations research.  

Simulated annealing (SA) consists of a Markov Chain Monte Carlo algorithm (MCMC) that samples the Gibbs distribution together with an annealing schedule.  The temperature parameter of the MCMC is adjusted during the run of the algorithm according to the annealing schedule which starts at some high temperature where the MCMC mixes rapidly to some low target temperature where mixing is slow. In optimization applications, the target temperature is chosen so that there is a significant probability of finding an optimal or near optimal solution (ground state).  The hope, born out by both theory and numerical experiments, is that for finding optima it is much more efficient to anneal to a low temperature than to simply run the MCMC at the low temperature.

Simulated annealing was introduced to find optima (ground states) but it can be extended \cite{Neal2001,Jarz97} to be used as a Gibbs sampler.  In sampling mode, it is known as annealing importance sampling (\AIS) and consists of multiple independent runs of SA.  Gibbs averages are obtained as weighted averages over these runs.   Population annealing (\PA) consists of many parallel, interacting runs of SA.  At each step in the anneal, a birth-death process, also known as resampling, removes some of these runs and copies others to keep the population near the equilibrium distribution and \PA natively functions as a Gibbs sampler.
A similar idea is implemented in diffusion Monte Carlo \cite{ReToGo90} and substochastic Monte Carlo \cite{JaLa17} where the population represents a wave function and resampling implements time evolution.

In this paper, we explicitly discuss the optimization of annealing schedules for population annealing and annealed importance sampling although the ideas are expected to carry over, at least qualitatively, to other annealing algorithms.  Our formalism specifies annealing schedules that, for a given amount of computational work, minimize systematic errors in sampling the Gibbs distribution.  The same schedules, with end points at sufficiently low temperatures, are expected to work well for optimization.  

A great deal of work has been done to determine optimal annealing schedules for SA.  It can be proved  \cite{GeGe84,Hajek88} that \SA will converge to the global optimum if the temperature schedule decreases to zero logarithmically slowly but these results are not useful in practical application.  A variety of ad hoc schedules have been proposed and studied, see, for example, \cite{cohn1999simulated,nikolaev2010simulated} and references therein.  The proposed annealing schedules that are most closely related to the present work are based on constant ``thermodynamic speed" \cite{SaNu88,AnGo94}.  Less effort has been devoted to optimizing annealing schedules for \PA with several proposals discussed in Refs.\ \cite{BaWeBoJaSh17,BaPaWaKa18,AmMa18}.

Annealing algorithms are most frequently deployed with a single control parameter, usually inverse temperature, however, there are many applications where multiple control parameters are useful.  In statistical physics, examples include sampling Gibbs distributions with external fields or chemical potentials.  In optimization applications, parameters are employed to enforce constraints via penalty terms.  During the anneal these penalties are typically increased so that constraints are initially soft but fully enforced at the end of the anneal. In the present work we develop a formalism for optimized annealing schedules for multiple control parameters.

The paper is organized as follows.  In Sec.\ \ref{sec:PA} we review population annealing and develop the theoretical ideas for optimal annealing paths in the context of multiple control parameters and smooth annealing paths.  The new theoretical material is mostly contained in Secs.\ \ref{sec:contime} and \ref{sec:rhofPA}.  In \ref{sec:contime} we introduce  continuous time annealing and, in conjunction with Appendix \ref{app:limit}, describe a new resampling method, pairwise residual resampling, that can be carried out quasi-continously without introducing significant errors.  

The theory of optimal annealing paths is developed in Sec, \ref{sec:rhofPA} and is formally similar to the theory of optimal control of thermodynamic systems introduced by Sivak and Crooks \cite{SiCr12} and their collaborators \cite{ZuSiCrDe12}.  
Following Sivak and Crooks \cite{SiCr12} we observe that the optimal annealing schedules, at least near equilibrium, are geodesics relative to a metric defined by the so-called ``friction tensor".  The components of the friction tensor are integrals of time correlation functions of the forces conjugate to the control parameters.  In the familiar case of a single control parameter, usually inverse temperature, the conjugate force is the energy.  It is perhaps not surprising that optimizing annealing algorithms and optimizing the control of thermodynamics systems are closely related since annealing algorithms were motivated by thermodynamics annealing and are computer instantiations of controlled thermodynamic systems. 

The friction tensor formalism assumes the dynamics, here a Markov chain, is ergodic and near equilibrium throughout the annealing schedule.  However, in many realistic situations the annealing path passes through a region where ergodicity is broken, for example, at a first order transition or in the low temperature phase of spin glasses.  In Sec.\ \ref{sec:nonergodic} we extend the formalism to a simple case of ergodicity breaking and explore how  transitions into non-ergodic regions of parameter space contribute systematic errors in population annealing.  Within a non-ergodic region the global correlation times appearing in the standard friction tensor diverge and must be replaced by an average over correlation times within each isolated local minimum of the free energy, which  motivates the introduction of a restricted friction tensor in Sec.\ \ref{sec:rzeta}.  

In Sec.\ \ref{sec:ais}  we extend the ideas developed for population annealing to annealed importance sampling and discuss when one or the other algorithm is expected to be superior.  %In Sec. \ref{sec:results} 
%we present r the simulation methods in Sec.\ \ref{sec:methods}. 
In Sec.\ \ref{sec:simulations} we present results from population annealing simulations for finding spin glass ground states using various annealing schedules, These results show that the restricted friction tensor schedule is superior to two of the commonly used schedules for population annealing.  The paper closes with a discussion in Sec.\ \ref{sec:conclude}.

\section{Population Annealing}
\label{sec:PA}
Population Annealing is a Sequential Monte Carlo method \cite{DeDoJa06} that sequentially transforms or ``anneals'' a population of $R$ replicas of the system  from an easy to sample region of parameter space to a difficult to sample region of parameter space.  
During the anneal, two processes are used to carry out the transformation of the population from one distribution to another.  The first is the action of a Markov chain Monte Carlo (MCMC) algorithm whose invariant measure is the Gibbs distribution  at the current value of the parameters. The second step consists of resampling the population, described in detail below.  Simulated Annealing can be understood as a special case of \PA where the population has one member ($R=1$) and, therefore, no resampling occurs.

Annealing is carried out following an annealing schedule, which is a trajectory in the space of parameters of the Gibbs distribution.  In many applications there is a single parameter, the inverse temperature $\beta$, which controls the average energy of the Gibbs distribution.  In this work, we consider annealing in multi-dimensional parameter spaces.  

We write the generalized Gibbs distribution, $\pi(\x,\lamb)$ for the probability of microstate $\x$ given the vector of parameters $\lamb= \{\lambda_i \}$ in the form 
\begin{equation}
\label{eq:gibbs}
    \pi(\x,\lamb) = \exp (F(\lamb) - E(\x,\lamb)),
\end{equation}
where $E(\x,\lamb)$ is the dimensionless energy as a function of system configuration $\x$ and $ F(\lamb)$ is the dimensionless free energy defined as,
\begin{equation}
    F(\lamb)=-\log \sum_\x \exp(- E(\x,\lamb)).
\end{equation}
We use the notation $\langle A \rangle_{\lamb}$ to indicate an average of observable $\obs$ over the generalized Gibbs distribution with control parameter vector $\lamb$.

In many cases, the parameters appear linearly in the energy function.  For example, consider an Ising model with coupling strength $J$ in an external field $h$. The dimensionless energy function is given by, 
\begin{equation}
\label{eq:ising}
    E(\{S_i\},\{\beta J,\beta h\}) = -\beta J\sum_{\langle i,j \rangle} S_i S_j - \beta h \sum_i S_i 
\end{equation}
where the spins $S_i$ take values $\pm 1$ and the summation is over the edges of the graph of coupled spins.  The two control parameters are $\lambda_1=\beta J$ and $\lambda_2=\beta h$.

The annealing schedule, $\lamb(t)$ is parameterized by time $t$, which ranges over the interval $0$ to $T$.   Time is measured in units of elementary moves of the MCMC method.  For example, in the case of the Ising model with the Metropolis algorithm as the MCMC a unit of time corresponds to attempting a single spin flip.  
In standard implementations of \PA and \SA time progresses in discrete steps but, to simplify the analysis, here we consider time to be a real parameter and assume that elementary moves of the MCMC occur as a Poisson process, as in the case of Glauber dynamics.  An elementary  MCMC move at time $t$ is governed by parameters $\lamb(t)$. In the case of \PA, time is considered to run in parallel for each member of the population so the annealing time variable is independent of the population size $R$.

\subsection{Resampling and Free Energy}
For \PA, values of the control parameters are taken to be stepwise constant functions of $t$ with jumps at regular intervals separated by step time $dt$.  Resampling is taken to occur instantaneously at the time of the steps in $\lamb(t)$ and is a birth-death process in the population of replicas.  Let $t$ be a resampling time and $t^+$ be a time immediately after the parameter change at time $t$ (more mathematically, $t^+$ is shorthand for taking a one-sided limit).  Let $\wt_r(t)$ be the reweighting factor for replica $r$ between the Gibbs distribution with parameters $\lamb(t)$ and $\lamb(t^+)$, 
\begin{align}
        \label{eq:tau}
        \wt_r(t) =\frac{1}{Q(t)} e^{-E(\x_r(t), \, \lamb(t^+)) + E(\x_r(t),\, \lamb(t))},
        \end{align}
where $\x_r(t)$ is the state of replica $r$, $R$ is the population size, and $Q(t)$ is a normalization factor used to control the population size,
        \begin{align}
        \label{eq:Q}
        Q(t) = \frac{1}{R}\sum_{r=1}^{R} e^{-E(\x_r(t) , \, \lamb(t^+)) + E(\x_r(t) ,\, \lamb(t))}.
\end{align}
The reweighting factors determine the (joint) distribution of the copy numbers $n_r$, the number of copies in the population at time $t^+$ of replica $r$ at time $t$. The copy numbers are non-negative integers whose expectations satisfy, \begin{equation}
\label{eq:ntau}
\E(n_r)=\wt_r.    
\end{equation}
Note that $n_r=0$ means replica $r$ is removed from the population and $n_r=2$ means a copy of the replica is added to the population.  Equation \eqref{eq:ntau} ensures that if the population at time $t$ is a good sample from the Gibbs distribution with parameters $\lamb(t)$ then, after resampling, the population at time $t^+$ will be a good sample from the Gibbs distribution with parameters $\lamb(t^+)$.

There are many ways to implement resampling that satisfy Eq.\ \eqref{eq:ntau} with either fixed population size  or variable population size  \cite{DoCa05,gessert2023resampling}.  Population annealing is typically implemented with nearest integer resampling \cite{WaMaKa15b,gessert2023resampling}.   In nearest integer resampling $n_r$ is independently chosen from either the floor or ceiling of $\wt_r$ such that
    Prob$(n_r=\lfloor \wt_r \rfloor)=(R/R(t))(\lceil \wt_r \rceil-\wt_r)$, and Prob$(n_r=\lceil \wt_r \rceil)=(R/R(t))(\wt_r-\lfloor \wt_r \rfloor)$ 
where, $R(t)$ is the population size at time $t$ before the resampling event and the factor $(R/R(t))$ is employed so that on each step the population is pushed toward the target size, $R$.  The formalism developed below assumes resampling with a fixed population size, which is discussed in  detail in the Appendix.  

At any time during the anneal, an estimator for observable $\obs$ can be obtained from the population average, $\tilde{\obs}$,
\begin{equation}
\label{eq:At}
    \tilde{\obs}(t) = \frac{1}{R} \sum_r \obs_r(t),
\end{equation}
where $\obs_r(t)$ is the value of $\obs$ in replica $r$ at time $t$.  

The resampling normalization, $Q(t)$ [Eq.\ \eqref{eq:Q}] provides an estimator of the free energy $\fe$ acccording to  
    \begin{equation}
    \label{eq:discreteF}
    \fe(\lamb(t_2))=F(\lamb(t_1))-\sum_{t_1 < t \leq t_2}  \log Q(t) 
\end{equation}
where the sum is over resampling times in the interval $(t_1,t_2]$.
This relation follows from the Jarzinski equality \cite{Jarz97a,Jarz97} and is also derived in the context of population annealing in Ref.\ \cite{Mac10a}.

\subsection{Continuous Time Annealing}
\label{sec:contime}
Although practical applications of annealing algorithms employ piecewise constant schedules, here we will develop a formalism for continuous annealing schedules.  We can achieve arbitrarily good approximations to a smooth $\lamb(t)$ by successively subdividing the time step, $dt$ between resampling in a piecewise constant schedule.  Henceforth, in this work we assume the schedule is a smooth function of time.  
It is not obvious that resampling is well-behaved when the time, $dt$ between resamplings becomes small.  In the Appendix we introduce a new resampling scheme, which we call pairwise residual resampling (\prr ), that is well-suited to doing frequent resampling.  Pairwise residual resampling shares features with nearest integer resampling, systematic resampling and residual resampling \cite{gessert2023resampling} but is designed to work only in the situation that the weights $\wt_r$ are very close to unity as is the case for sufficiently small time steps.  Other resampling schemes such as multinomial and Poisson resampling are not well behaved in the $dt \rightarrow 0$ limit.

For small time steps, the resampling normalization $Q(t)$ can be expanded to leading order in $dt$ as,
\begin{align}
\label{eq:Qexpand}
    Q(t) &= 1-\frac{1}{R}\sum_{r=1}^{R} E(\x_r(t), \, \lamb(t^+)) - E(\x_r(t) ,\, \lamb(t))\\
    \label{eq:Qlimit}
    &\rightarrow 1-\frac{dt}{R}\sum_{r=1}^{R} \dot{ \lambda}_i \frac{\partial E_r(\lamb)}{\partial \lambda_i},
\end{align}
where $E_r$ is shorthand for the dimensionless energy of replica $r$ and summation over repeated indices is assumed. The expected number of copies of replica $r$ reduces to 
\begin{equation}
\label{eq:taudt}
    \wt_r \rightarrow 1- dt \dot{ \lambda}_i (X_i^r - \tilde{X}_i).
\end{equation}
Here $X^r_i$ is the $i^{\rm th}$ conjugate force of the $r^{\rm th}$ replica,
\begin{equation}
\label{eq:X}
    X^r_i = -  \frac{\partial E_r(\lamb)}{\partial \lambda_i}
\end{equation}
and $\Xt_i$ is the population average of $X^r_i$ (see Eq.\ \eqref{eq:At}).  In the case of the Ising model in a field, Eq.\ \eqref{eq:ising},  $X_1=\sum_{\langle i,j \rangle} S_i S_j$ is the dimensionless coupling energy and $X_2=  \sum_i S_i$ is the dimensionless magnetization.

From Eqs.\ \eqref{eq:discreteF}, \eqref{eq:Qlimit} and \eqref{eq:X} in the $dt \rightarrow 0$ limit, the free energy estimator  becomes,
\begin{equation}
\label{eq:F}
   \fe(\lamb(t_2)) =F(\lamb(t_1)) + \int_{t_1}^{t_2} dt^\prime\; \dot{ \lambda}_i \tilde{X}_i(t)  ,
\end{equation}
The integral in Eq.\ (\ref{eq:F}) represents the generalized work associated with the anneal from $t_1$ to $t_2$.

We conclude this section introducing some notation used in the rest of the paper.  The average of a quantity $A$  over the Gibbs distribution with  parameters $\lamb$  is written as $\langle A \rangle_{\lamb}$. If we carry out many runs of \PA with schedule $\Lamb= \{\lamb(t): 0\leq t \leq T\}$ then the average over runs of the estimator $\tilde{A}$ from each run is written as $\langle \tilde{A}(t) \rangle_{\Lamb}$.  Note $\tilde{A}(t)$ may depend on the annealing path up to time $t$.  On the other hand, if the anneal is sufficiently slow so that the population remains close to equilibrium then $\langle \tilde{A}(t) \rangle_{\Lamb} \approx \langle A \rangle_{\lamb(t)}$.  For example, the \PA average for the free energy is
\begin{equation}
\label{eq:Fave}
   \langle \fe(\lamb(t_2)) \rangle_{\Lamb} =F(\lamb(t_1)) + \int_{t_1}^{t_2} dt^\prime\; \dot{ \lambda}_i \langle \tilde{X}_i(t) \rangle_{\Lamb} ,
\end{equation}
and, for an infinitely slow anneal we recover the standard result for thermodynamic integration, 
\begin{equation}
\label{eq:Fexact}
   F(\lamb(t_2))  =F(\lamb(t_1)) + \int_{t_1}^{t_2} dt^\prime\; \dot{ \lambda}_i \langle X_i(t) \rangle_{\lamb(t)} ,
\end{equation}

\subsection{Culling Fraction}
\label{sec:cull}

An important metric associated with resampling is the culling fraction $\epsilon$, the expected number of replicas removed from the population during resampling.  For both nearest integer resampling and pairwise residual resampling (see the Appendix), it straightforward to see \cite{AmMa18} that the culling fraction for a single resampling step is,  
\begin{equation}
\label{eq:eps}
    \epsilon = \frac{1}{R}\sum_{\{r| \wt_r<1 \} } (1-\wt_r).
\end{equation}
Note that the complement, $1-\epsilon$,  can also be interpreted as the overlap between the Gibbs distributions before and after a nearest integer resampling step. Two quite effective proposed annealing schedules with a single control parameter for \PA are a fixed culling fraction schedule \cite{AmMa18} and, equivalently for nearest integer resampling, a fixed overlap schedule  \cite{BaWeBoJaSh17}.  In Sec.\ \ref{sec:results} we compare the friction tensor schedules proposed below to fixed culling fraction schedules.

In the $dt \rightarrow 0$ limit all $\wt$'s are near one, so the culling fraction for a single resampling step is infinitesimal. The cumulative culling fraction $\epsilon(t_1,t_2)$ over  segment $(t_1,t_2)$ of the annealing path can be obtained by expanding $\wt_r$ to leading order in $dt$ and integrating over the annealing path segment.  Furthermore,  since  the population average of $\wt_r$ is one (or very nearly one for nearest integer resampling), the sum in Eq.\ \eqref{eq:eps} can be equivalently expressed as being over replicas with $\wt_r > 1$ except that the sign of the summand is changed. Using Eq.\ \eqref{eq:taudt} we obtain an expression for the cumulative culling fraction along an annealing segment,
\begin{equation}
\label{eq:eps12}
   \epsilon(t_1,t_2) =  \frac{1}{2R}\int_{t_1}^{t_2} dt   \langle \sum_r |\dot{ \lambda}_i(X^r_i -\Xt_i)| \rangle_{\Lamb}.
\end{equation}
If the population remains close to equilibrium, the nonequilibrium average, $\langle \cdots \rangle_{\Lamb}$ can be replaced by an equilibrium average $\langle \cdots \rangle_{\lamb(t)}$,
\begin{equation}
   \label{eq:eps12eq}
   \epsilon(t_1,t_2) =  \frac{1}{2}\int_{t_1}^{t_2} dt^\prime\; \langle | \dot{ \lambda}_i \delta X_i | \rangle_{\lamb(t)} . 
\end{equation}
In this expression, $\delta X = X - \langle X \rangle_{\lamb(t)}$.

If furthermore, the Gibbs distribution is a multivariate Gaussian, as is often a good approximation, then we have,
\begin{equation}
   \label{eq:eps12gaussian}
   \epsilon(t_1,t_2) = \int_{t_1}^{t_2} dt^\prime\; \sqrt{\frac{ \dot{ \lambda}_i \sigma_{ij}^2 \dot{ \lambda}_j}{2 \pi}}, 
\end{equation}
where $\sigma_{ij}=\langle \delta X_i \delta X_j \rangle_{\lamb(t)}$ is the covariance matrix. In the simple case that the only control parameter is inverse temperature the integrand reduces to $\dot{\beta}\sigma_{e} /\sqrt{2 \pi}$ where $\sigma_e$ is the variance of the dimensionless energy.

\subsection{Weighted averaging and systematic errors}

A simple average over multiple independent runs of \PA reduces statistical but not systematic errors. 
Multiple independent runs of \PA can be combined using weighted averaging to improve both statistical and systematic errors in measuring either the free energy or an observable \cite{WaMaKa15b}.  Let $\tilde{\obs}_m$ be the estimator of observable $\obs$ from run $m$ of PA and let $\tilde{F}_m$ be the estimator of the dimensionless free energy in run $m$ of PA. The weighted average over $M$ runs, $\overline{\obs}$ is given by,
\begin{equation}
\label{eq:weightA}
  \overline{\obs}= \frac{1}{M} \sum_{m=1}^M \tilde{\obs}_m \exp(-\fe_m+ \overline{F}), 
\end{equation}
where $\overline{F}$ is the weighted average free energy,
\begin{equation}
\label{eq:weightF}
   \overline{F}= - \log \frac{1}{M}\sum_{m=1}^M  \exp(-\fe_m).
\end{equation}
It is assumed in these expressions that the  population size $R$ and the annealing schedule in all runs are the same.  For fixed $R$, in the limit of $M \rightarrow \infty$, the weighted free energy and all observables converge to their exact Gibbs value.  In optimization applications, weighted averaging corresponds simply to choosing the best solution from all runs.

A single run of \PA with finite population size will suffer from systematic errors.  The magnitude of these errors can be obtained by comparing ordinary averaging over runs, which does not suppress systematic errors to weighted averaging, which becomes exact for large $M$. The central limit theorem suggests that for large $R$, the estimators for observables and the free energy should become Gaussian and then the systematic error is related to the (co)variances according to \cite{WaMaKa15b},
\begin{equation}
    \Delta \tilde{\obs} = \mathrm{cov}(\tilde{\obs},\fe)
\end{equation}
and 
\begin{equation}
    \Delta \fe = \mathrm{var}(\fe)/2,
\end{equation}
where $ \Delta \tilde{\obs}$ is the difference between the exact value of the observable and the expected result from a single \PA run,
\begin{equation}
 \Delta \tilde{\obs} = \langle \tilde{\obs} \rangle_{\Lamb} - \langle \obs \rangle_{\lamb},  
\end{equation}
and $\langle \cdots \rangle_{\Lamb}$ is an (ordinary) average over runs of \PA (or SA) with a schedule $\Lamb$ that terminates at control parameter values $\lamb$.  The analogous definition of $\Delta \fe$ is
\begin{equation}
 \Delta \fe = \langle \fe \rangle_{\Lamb} - F(\lamb),  
\end{equation}
where $F(\lamb)$ is the free energy associated with the Gibbs distribution at control parameter values $\lamb$.

Inspired by the central limit theorem,  we assume that for large $R$ both $\mathrm{cov}(\tilde{\obs},\fe)$ and  $\mathrm{var}(\fe)$ scale as $1/R$ and we can then re-write the systematic errors as
\begin{equation}
    \Delta \fe = \rho_f /2R,
\end{equation}
and 
\begin{equation}
    \Delta \tilde{\obs} = \rho_f  C(\obs) /R
\end{equation}
where
\begin{equation}
\label{eq:rhof}
    \rho_f = \lim_{R \rightarrow \infty} R \, \mathrm{var}(\fe)
\end{equation}
and
\begin{equation}
     C(\obs) = \lim_{R \rightarrow \infty}  \frac{\mathrm{cov}(\tilde{\obs},\fe)}{\mathrm{var}(\fe)}
\end{equation}
The quantity $\rho_f$, introduced in the context of \PA in \cite{WaMaKa15b}, sets the scale of systematic errors in free energy measurements.  Furthermore, if the coefficient $C(A)$ is approximately independent of the annealing schedule then $\rho_f$ also sets the scale of systematic errors in the observable $A$.  Since $A$ is calculated from the final population and not from the whole trajectory, we expect the $C(A)$ is only weakly dependent on the annealing path.

Thus $\rho_f$ can be loosely interpreted as the the population size required to reach equilibrium. In order to minimize the computational work of doing equilibrium sampling with \PA the annealing schedule should be chosen to minimize $\rho_f$.  Much of the remainder of the paper is concerned with obtaining tractable expressions for $\rho_f$ that are amenable to minimization.

While $\rho_f$ is related to systematic errors, other measures have been proposed to estimate statistical errors in PA. One of these, $\rho_t$ (see Refs.\ \cite{WaMaKa15b,AmMa18,gessert2023resampling})  is based on the distribution of family sizes.  In this context a family is a set of replicas descended from a single replica in the initial population and $\rho_t$ is the sum of the squares of the family sizes divided by the population size.  Thus, at the beginning of the simulation, $\rho_t=1$ and is bounded by $\rho_t \leq R$.  As shown in \cite{AmMa18}, for the Ising spin glass $\rho_t$ and $\rho_f$  typically differ by a constant that is small compared to $\rho_f$ itself.  Because of this close connection and the ease of measuring $\rho_t$, we use this measure to characterize the hardness of a spin glass samples studied in Sec.\ \ref{sec:results}.  A second measure of statistical errors is the effective population size introduced in \cite{WeBaShJa21}, which is bounded above by $R/\rho_t$.

\subsection{Friction Tensor}
\label{sec:rhofPA}
In this section we obtain a formal expression for $\rho_f$ and show that in the near equilibrium approximation, it can be expressed in terms of the friction tensor introduced in Ref.\ \cite{SiCr12}.  The variance of the free energy estimator is obtained by squaring Eq.\ \eqref{eq:F} and subtracting mean values so that Eq.\ \eqref{eq:rhof} becomes,
\begin{equation}
\label{eq:varF}
   \rho_f= %\nonumber \\
   \lim_{R \rightarrow \infty} R \int_{0}^{T} dt^\prime\int_{0}^{T} dt''\; \dot{ \lambda}_i(t') \langle \delta\tilde{X}_i(t') \delta\tilde{X}_j(t'')\rangle_{\Lamb} \dot{ \lambda}_j(t'')  ,
\end{equation}
where  $\delta A = A - \langle A \rangle_{\Lamb}$.  This expression is a time integral over a matrix of non-equilibrium correlation functions of the conjugate forces.  The dynamics in the non-equilibrium average includes both the dynamics of the MCMC and of resampling, both of which promote the decay of correlations. In the more general setting of sequential Monte Carlo algorithms, similar expressions for the variance of the free energy estimator have been developed~\cite{DeDoJa06,chopin2004}.

To gain greater insights into the meaning of Eq.\ \eqref{eq:varF} and to obtain a more tractable expression we make two related approximations:
\begin{enumerate}
    \item The population remains close to an equilibrium sample during the anneal.  
    \item  The control parameters vary slowly compared to the time scale for the decay of correlations. 
\end{enumerate}
Given these approximations we set $t'=t''$ in the time argument of the parameter velocities and we replace the non-equilibrium average by an equilibrium average at the current values of the control parameters.  The resulting expression takes the form of a time integral over a bilinear form,
\begin{equation}
\label{eq:rhofzeta}
    \rho_f = \int_0^{T} d t \dot{ \lambda}_i(t)  \zeta_{ij}(\lamb(\tau))  \dot{ \lambda}_j(t),
\end{equation}
where $\zeta_{ij}(\lamb(\tau))$ is the `friction tensor',
%JM add citation for friction tensor
\begin{equation}
\label{eq:zeta}
 \zeta_{ij}(\lamb(t))= 2 \int_{0}^{\infty} dt'\langle\delta X_i(0) \delta X_j(t') \rangle_{\lamb(t)}.  
\end{equation}
  The assumption that correlations decay quickly justifies taking the limit to infinity in the integral defining the friction tensor. The friction tensor can be further decomposed into the outer product of the matrix of equilibrium correlation times and equal time covariances,
\begin{equation}
    \zeta_{ij} = 2 \tau_{ij}\sigma_{ij}^2,
\end{equation}
where $\sigma_{ij}= \langle\delta X_i \delta X_j \rangle_{\lamb(t)}$ and 
\begin{equation}
    \tau_{ij} =   \int_{0}^{\infty} dt'\frac{\langle\delta X_i(0) \delta X_j(t') \rangle_{\lamb(t)}}{ \langle\delta X_i \delta X_j \rangle_{\lamb(t)}}.
    \label{eq:integrated-autocorrelation}
\end{equation}
In the friction tensor approximation, the requirements for a good annealing path are intuitively plausible: regions of parameter space with large values of variances or autocorrelation times should be avoided or, if they cannot be avoided, traversed slowly.

From this point onward  there are two directions to explore.  The first is to obtain a more detailed understanding of the implications of Eq.\ \eqref{eq:rhofzeta}.  Fortunately, this equation is formally identical and also conceptually related to the expression obtained by Sivak and Crooks \cite{SiCr12,ZuSiCrDe12} for minimum excess work paths in nonequilibrium thermodynamics so we can take over their analysis with minimal modification.
 
The second, more difficult task is to understand the role of resampling.  In going from Eq.\ \eqref{eq:varF} to Eq.\ \eqref{eq:rhofzeta} resampling has been explicitly removed from the expression since the dynamics in the definition of the friction tensor is at a single temperature and resampling only plays a role when the temperature changes. It appears we have thrown away the baby with the bath water.

\subsubsection{Optimal Annealing Paths for Population Annealing}
\label{sec:optanneal}
The friction tensor formulation of $\rho_f$ in Eqs.\ \eqref{eq:rhofzeta} and \eqref{eq:zeta}  is formally identical to the expression found by Sivak and Crooks \cite{SiCr12} for minimum excess thermodynamic work, which allows us to adopt their analysis directly. They showed that the friction tensor defines a pseudo-metric space whose geodesics are minimum dissipation paths in a space of thermodynamic parameters.  In one dimension, the geodesic simply determines the  speed along the path and, when translated into an annealing schedule in the single parameter of inverse temperature,  $\beta$ the result is (\cite{SiCr12}, Eq.\ 12),
\begin{equation}
\label{eq:dotbeta}
    \dot{\beta} \propto \zeta^{-1/2} , 
\end{equation}
where, in this case,  $\zeta = 2 \tau_e C/\beta^2$, $C =\beta^2 \sigma^2_e$, is the heat capacity, $\sigma_e$ is the standard deviation of the energy and $\tau_e$ is the integrated autocorrelation time of the energy.   This annealing schedule has the property that the rate of increase in $\rho_f$ is constant along the path. 

The annealing schedule \eqref{eq:dotbeta} can be re-written as $\dot{\beta} \propto 
\beta/ \sqrt{\tau C}$, which makes clear the similarity to the constant thermodynamic speed schedule for \SA proposed in Refs.\ \cite{SaNu88,AnGo94}, which takes the form $\dot{\beta} \propto \beta/\tau \sqrt{C}$.  Here $\tau$ is the `apparent relaxation time' of the energy. Assuming that `apparent relaxation time' is interpreted as the energy integrated autocorrelation time the only difference between these schedules is an extra factor of $\sqrt{\tau_e}$ in the denominator.  

Rotskoff and Crooks \cite{RoCr15} show how to compute geodesics in the two-dimensional space of coupling energy and magnetic field for the Ising model and these ideas could be applied to determining annealing paths in the temperature-field plane.

\subsubsection{Ergodicity Breaking and the  Role of Resampling}
\label{sec:nonergodic}

The friction tensor formalism developed at the end of Sec.\ \ref{sec:rhofPA} does not take into account resampling since it uses Gibbs averages taken at a single value of the control parameters. In this section we discuss the role of resampling. In regions of parameter space where correlation times are short, the role of resampling is relatively limited and difficult to quantify.  On the one hand, resampling helps keep the population close to the current Gibbs distribution so that the local equilibrium  approximation in going from the full expression, Eq.\ \eqref{eq:varF} to the friction tensor formulation, Eq.\ \eqref{eq:rhofzeta} is more accurate.  On the other hand, resampling increases the variance of the free energy and covariance of the free energy with observables because it is a stochastic process.  It is not clear which of these effects dominates for a given problem so that it may be that AIS or simply using the MCMC at the target values of the control parameters is more efficient.

Resampling becomes important in regions of parameter space where  correlation times become large so that, by itself, the MCMC cannot reach equilibrium. Here resampling may be required to maintain equilibrium as the control parameters change.  While a full analysis of the combined effects of the resampling and the MCMC is beyond the scope of this work, we will investigate a simple and relatively generic situation to show how the friction tensor formulation can be modified to include both non-ergodicity of the MCMC  and the contributions of resampling to $\rho_f$.

In the simplified model, there are two disconnected regions in configuration space below an ergodicity breaking transition.  Within each region the MCMC is ergodic but transitions between regions do not occur on the time scales of the simulation.  
We assume that within each of the two regions,  the MCMC is able to achieve local equilibrium in the region on a reasonable time scale and it is these times that will enter into the modified friction tensor formulation.  To further simplify the problem, suppose that inverse temperature $\beta$ is the only control parameter,  energy $E$ is its conjugate field and the only relevant time scales are the energy autocorrelation times in the two regions. Finally, we assume that the difference between the entropies and energies of the two regions are independent of $\beta$. Let $\de >0$ be the average energy difference. The entropy difference can be parameterized by a transition temperature, $\beta_c$ where the two regions are equally probable in the Gibbs distribution.  The conclusions we obtain for this simple two region model should apply, at least qualitatively, to situations where configuration space fractures into multiple disconnected regions with more complex free energy landscapes.
 %JM maybe should have the even simpler assumption of constant energy and entropy
 
 The simple two-region model is discussed in Section IV of Ref.\ \cite{Mac09}.  The key result is that the probability, $p(\beta)$ of being in the lower energy (and lower entropy) regions obeys a Fermi distribution,
\begin{equation}
    p(\beta) = \frac{1}{1+ e^{-(\beta-\beta_c)\de}}
\end{equation}
and that the entropy difference is $\beta_c \de$.  Suppose that the MCMC becomes effectively non-ergodic at $\bn$. The interesting and computationally difficult situation is when non-ergodicity occurs before the transition, $\bn<\beta_c$, while the target temperature is after the transition,   $\beta(T)> \beta_c$ so that the low energy region dominates the distribution at the target temperature but is improbable at the ergodicity transition.  As the anneal proceeds beyond $\bn$, resampling increases the fraction of the population in the low energy region but stochasticity in the number of replicas in the low energy region at $\bn$ is not eliminated until well beyond $\beta_c$  when nearly all the population is in the low energy region.

To evaluate the contribution to $\rho_f$ from resampling in the two-region model, suppose, without loss of generality, the energy in low energy region is zero, 
%JM perhaps make this assumption earlier
and let $\rl$ be the fraction of replicas  in this zero energy region.  Since the annealing schedule is in $\beta$ the only conjugate field is the energy and $\Xt = (1- \rl) \de$.

If we ignore the stochasticity of resampling, then from Eq.\ \eqref{eq:tau}, we have the following differential equation for the population in the low energy region as a function of $\beta$,
\begin{equation}
    \frac{d \rl}{d \beta} = \de \rl ( 1- \rl),
\end{equation}
whose solution with initial condition, $\rl=\rln$ is,
\begin{equation}
\label{eq:rl}
    \rl(\beta) = \frac{\rln e^{(\beta-\bn) \de}}{1+\rln(e^{(\beta-\bn)\de}-1)}.
\end{equation}

Since we are only concerned with resampling, we use $\beta$ as the independent variable instead of time in calculating the double integrals defining $\rho_f$ [Eq.\ \ref{eq:varF}] and obtain, for an anneal from the ergodicity transition to infinite $\beta$, 
\begin{equation}
\label{eq:rhof2}
    \rho_f= \lim_{R \rightarrow \infty} \de^2 R \int_{\bn}^{\infty} d\beta^\prime\int_{\bn}^{\infty} d \beta''\; \langle \delta\rl(\beta') \delta\rl(\beta'')\rangle_{\Lamb},
\end{equation}
where $\langle \cdots \rangle_{\Lamb}$ is here an average over Gaussian initial conditions.
 At the ergodicity transition the population size in the zero energy region, $R \rl(\bn)$, is a Poisson random variable with mean and variance $R p(\bn)$.  For $R p(\bn) \gg 1$ the distribution of $\rln$ approaches a sharply peaked Gaussian distribution with mean $p(\bn)$ and variance $p(\bn)/R$. 

In order to evaluate Eq.\ \eqref{eq:rhof2} we switch the order of averaging and integration.  The only integral to be done is 
\begin{eqnarray}
    \label{eq:gx}
     g(x) = \int_{0}^{\infty} dz \; \frac{x e^{z \de}}{1+x(e^{z\de}-1)}-1 \\
     =\log(x)/\de.
\end{eqnarray}
The first term in the integrand is transcribed from Eq.\ \eqref{eq:rl}  and the one is subtracted to make the integral convergent but does not affect the final result for $\rho_f$.

In terms of $g$, the expression for $\rho_f$ takes the form,
\begin{equation}
\label{eq:rhofg}
    \rho_f= \lim_{R \rightarrow \infty} \de^2 R \left[ \langle g(\rln)^2\rangle_{\Lamb}- \langle g(\rln)\rangle^2_{\Lamb} \right].
\end{equation}
To carry out these Gaussian averages in the large $R$ limit, both $g(x)$ and $g(x)^2$ are expanded in a Taylor series around the mean $\rln=p(\bn)$ and terms to second order in $(\rln-p(\bn))$ are kept.  The final result is  simple and intuitive,
\begin{equation}
\rho_f = 1/p(\bn).
\end{equation}
The intuition behind this result is that once the population size exceeds $1/p(\bn)$ it becomes likely that the low energy state is in the population and, via resampling, can grow to take over the population.  Beyond this point, systematic errors are suppressed proportional $1/R p(\bn)$.

It should be noted that  a subtle error has been made in going from the exact Poisson distribution for the population in the low energy region to the approximate Gaussian distribution. Specifically, with probability $e^{-R p(\bn)}$, there are {\em no} members of the population in the low energy region when the system becomes non-ergodic.  In this situation,  the algorithm completely fails.  Thus is it is more correct to say that $\rho_f$ as calculated above controls systematic errors when the failure probability is small, $e^{-R p(\bn)} \ll 1$.

Although the above calculation applies exactly to the simple two-region model, the qualitative result should hold if there are multiple disconnected competing regions, if there are first-order-like transition preceded by ergodicity breaking among several regions in configuration space along an annealing trajectory in several dimensions.  Specifically, upon crossing each first order like transition one expects a contribution to $\rho_f$ that behaves as $1/p$ where $p$ is the probability of being in the lower energy region when ergodicity is broken. 

Thus far we have assumed the MCMC is capable of equilibrating replicas within each disconnected region but we have not quantified the contribution to $\rho_f$ associated with the MCMC due to correlation times and equal time covariance.   To estimate these contributions it is necessary to modify the friction tensor formulation because resampling takes care of redistributing replicas between disconnected regions and the MCMC only decorrelates replicas within these regions.  In the next section we introduce a restricted friction tensor for use in non-ergodic regions.   

\subsubsection{Restricted friction tensor}
\label{sec:rzeta}
Suppose there are $k$ disconnected regions in configuration space and that as a function of time along the annealing path, the equilibrium probability of being in each region $\alpha$ is $p_\alpha(\lamb(t))$.  We assume that resampling distributes the population $\rl_\alpha(t) \approx  p_\alpha(\lamb(t))$ where $\rl_\alpha$ is the fraction of the population in region $\alpha$.  If we  ignore the contribution of resampling  to $\rho_f$ and proceed as before from Eq.\ \eqref{eq:varF} to Eqs.\ \eqref{eq:rhofzeta} and \eqref{eq:zeta} we obtain, for the restricted friction tensor, $\zeta^{\rm }_{ij}$
\begin{equation}
  \label{eq:zetar}
 \zeta^r_{ij}(\lamb)= 2 \int_{0}^{\infty} dt' \sum_\alpha p_\alpha(\lamb) \langle\delta_\alpha X_i(0) \delta_\alpha X_j(t') |\alpha \rangle_{\lamb},
\end{equation}
where $ \langle \cdots |\alpha \rangle_{\lamb}$ is a conditional Gibbs average over the (dynamically determined) regions labeled by $\alpha$ and $\delta_\alpha X = X -\langle X |\alpha \rangle_{\lamb}$.  Note that if one region dominates the average then this expression reduces to the unrestricted friction tensor expression.  When there are several important regions of configuration space, the restricted friction tensor is an average over the friction tensors associated with each region.

In general the conditional averages associated with the restricted friction tensor are not known, however, we can estimate the restricted friction tensor from the correlation functions appearing in the full friction tensor if there is a clear separation of time scales between the local relaxation time within regions and global relaxation time.  We have from the law of total covariance,
\begin{eqnarray}
\label{eq:lawcov}
  \langle\delta X_i(0) \delta X_j(t') \rangle =   \sum_\alpha p_\alpha \langle\delta_\alpha X_i(0) \delta_\alpha X_j(t') |\alpha \rangle  \\
  \nonumber
  +\sum_\alpha p_\alpha \left[\langle X_i \rangle - p_\alpha \langle X_i  |\alpha \rangle \right]\left[\langle X_j \rangle - p_\alpha \langle X_j  |\alpha \rangle \right].
\end{eqnarray}
Note the  second term on the RHS of Eq.\ \eqref{eq:lawcov} does not decay but if we subtract off the linearly increasing contributions to the integral defining the unrestricted friction tensor we obtain the restricted friction tensor.  In order to measure the restricted friction tensor, we can do a linear fit of the time integral whose limit is the full friction tensor,
\begin{equation}
   2 \int_{0}^{x} dt'\langle\delta X_i(0) \delta X_j(t') \rangle_{\lamb} \rightarrow  a_{ij}(\lamb) x  + \zeta^r_{ij}(\lamb), 
   \label{eq:linear-fit}
\end{equation}
where $a_{ij}(\lamb)$ the second term Eq.\ \eqref{eq:lawcov} is independent of $x$.  We can write the restricted friction tensor in the same form as Eq.\ \eqref{eq:zeta}, 
\begin{equation}
    \label{eq:rtau}
 \zeta^r_{ij}=2 \tau^r_{ij}\sigma^2_{ij}   
\end{equation}
where the restricted integrated correlation time matrix $\tau^r_{ij}$ is defined by this equation.
\section{Annealed Importance Sampling}
\label{sec:ais}
Simulated annealing  (\PA with $R=1$) can also be used with weighted averaging; this application of \SA is known as \textit{annealed importance sampling} (\AIS) \cite{Neal2001,Jarz97}. 
Annealing importance sampling permits simulated annealing to be used not only as an optimizer but also as a Gibbs sampler.    
In what follows, we will compare a single run of \PA with population $R$  to a run of \AIS consisting of the weighted average over $\Rs$ runs of \SA  to better understand the pros and cons of resampling.  In an abuse of notation, for \AIS we use $\fe$ and $\tilde{A}$ to represent the estimator resulting a single run of \AIS of size $\Rs$.  Multiple runs of \AIS with the same $\Rs$ can be additional combined using Eqs.\ \eqref{eq:weightA} and \eqref{eq:weightF}.
With this notation, combining Eq.\ \eqref{eq:F} and \eqref{eq:X} with $R=1$ and Eq.\ \eqref{eq:weightF} with $M=\Rs$, we have
\begin{equation}
    \label{eq:FSA}
   \fe(\lamb(t_2)) =F(\lamb(t_1)) -   
    \log \frac{1}{\Rs}\sum_{r=1}^{\Rs}  \exp\left(-\int_{t_1}^{t_2} dt^\prime\; \dot{ \lambda}_i X_i^r\right),
\end{equation}
where
\begin{equation}
    X_i^r=-\frac{\partial E_r(\lamb)}{\partial \lambda_i}.
\end{equation}
In \AIS, observables are obtained from the analog of Eq.\ \eqref{eq:weightA},
\begin{equation}
\label{eq:weightAIS}
  \tilde{\obs}= \frac{1}{\Rs} \sum_{r=1}^{\Rs} \obs_r \exp\left(-\int_{t_1}^{t_2} dt^\prime\; \dot{ \lambda}_i X_i^r+ \fe\right), 
\end{equation}
where $\obs_r$ is the value of the observable in run $r$ of SA.
The variance of $\fe$ for \AIS now involves the {\em exponential} of the integral along the annealing path.  If the variance of the exponential is itself small then the formula for $\rho_f$ for \AIS will reduce to the \PA formula except that two-time correlators are with respect to the dynamics of the MCMC alone and include no resampling. 

\subsection{The role of reweighting in Annealed Importance Sampling}
We can analyze the role of reweighting in \AIS in the context of the two-region model with the same assumptions as used in the previous sections.  The sum in Eq.\ \eqref{eq:FSA} contains two contributions, one from runs of \SA that find the low energy region and one from runs that do not so that 
\begin{equation}
    \fe = F_0 - \log\left( \rl(\bn) + (1-\rl(\bn)) e^{-(\beta-\bn) \de }\right)
\end{equation}
Thus $\rho_f$ for \AIS in the limit of large $\beta$ is given is simply,
\begin{equation}
    \rho_f = \lim_{\Rs \rightarrow \infty} \Rs \mathrm{var}(
    \log (\rl(\bn)) ),
\end{equation}
which gives exactly the same result as for \PA, $\rho_f = 1/p(\bn)$.

For the simple model of ergodicity breaking preceding a first-order-like transition, we see that AIS and PA perform equally well. However, for \PA we have ignored the stochasticity associated with resampling so, in principle, \AIS may be superior to \PA.  This conclusion conflicts with simulations that show that \PA with population size $R$ is far superior to $R$ runs of \SA for finding ground states of the three-dimensional Ising spin glass (Edward-Anderson model)  \cite{WaMaKa15}.  We believe this difference is related to the fact that spin glasses and perhaps many other systems undergo a sequence of several first-order-like transitions as the temperature is lowered.  This phenomenon is known as temperature chaos \cite{BrMo87,WaMaKa15a} in the spin glass literature.

Furthermore, let's suppose the landscape of these successive transitions is nested in the sense that it is easier for a replica that has already passed through one of these transitions to pass through the next transition.  In this situation \PA will be superior because resampling brings the population into the low-energy region of the first transition so that more replicas are exploring the subspace in which the next transition is likely to be found.  The simulation results of Ref.\ \cite{WaMaKa15} suggest that the landscape is indeed nested.

\section{Simulations}
\label{sec:simulations}

In this section, we describe the results of a computational study comparing the performance of several one-dimensional annealing schedules for population annealing applied to the  Ising spin glass.  To compare the performance of these schedules we measured both $\rho_f$ at low temperature and the probability of finding the ground state. 

\subsection{Numerical Methods}
\label{sec:methods}
We carried out PA simulations of 100 instances of the three-dimensional Ising spin glass with Gaussian disorder on a cubic lattice of size $10 \times 10 \times 10$ with periodic boundary conditions.  We simulated 100 disorder samples using 5 annealing schedules with a target population size of $R=10^5$ and nearest integer resampling.  These disorder samples had been previously studied in Ref.\ \cite{AmMa18}.  For each annealing schedule, we carried out two separate sets of simulations.  In the first set we allotted a total of 1000 Monte Carlo sweeps to each simulation and in the second set 3000 sweeps. In both cases, the elementary move was a Metropolis spin flip on a randomly chosen spin and one sweep consisted of 1000 elementary moves.  For each annealing schedule, the sweeps were divided among 300 temperature steps starting at $\beta=0$ and ending at $\beta=5$. For each schedule and each sweep number, we carried out 20 independent runs and estimated $\rho_f$ at $\beta=5$ from the variance of these runs. For the probability of finding the ground state for a given disorder realization, we used the fraction of the 20 runs that found the same ground state energy as found in Ref.\ \cite{AmMa18}.

We studied three friction tensor schedules satisfying $\dot{\beta} =1/\sqrt{\zeta}$ (see Sec.\ \ref{sec:optanneal} and Eq.\ \eqref{eq:dotbeta}), the commonly used fixed culling fraction schedule satisfying $\dot{\beta} =1/\epsilon$ (see Sec.\ \ref{sec:cull} ), and a constant $\beta$-step schedule, $\dot{\beta} =\mathrm{const}$.
The different annealing schedules are distinguished by the inverse temperature step size $\delta \beta$  and sweeps per temperature step $\delta s$ as a function of $\beta$.  In our continuous time formulation, it is only the ratio $\delta s/\delta \beta $ that is relevant to computing $\rho_f$ so that there are many ways to implement schedules for a given  $\dot{\beta}$. In the case of the friction tensor schedules, we choose three options that seem natural: (1) The number of sweeps per temperature step is constant, (2) the number of sweeps per temperature step yields a fixed culling fraction $\epsilon$ in each resampling step, and (3) the number sweeps is proportional to the energy integrated autocorrelation time, $\tau_e$.  If our simulations are close enough to the continuous time ($dt \rightarrow 0$) limit, there is expected to be little difference in the performance of these three schedules. 

\begin{table}[b!]
    \begin{center}
    \caption{
        Definitions of the annealing schedules investigated in this paper. The schedules are defined by the inverse temperature step size $\delta\beta$ and sweeps per temperature step $\delta s$ as a function of $\beta$. Note that the values given in this table are up to some normalization factors, which are used to fix the total number of temperature steps and sweeps. Here, $c_0$ is the uniform schedule, $\epsilon_0$ is the fixed culling schedule, and $\zeta_1,\zeta_2,\zeta_3$ are the three types of friction tensor schedules. $\epsilon$, $\zeta$, and $\tau_e$ are the culling fraction, the friction tensor, and the integrated autocorrelation time of the energy, respectively.}
        \label{table1: schedule definitions}
        \begin{tabular*}{\columnwidth}{@{\extracolsep{\fill}} l l l}
        \hline
        \hline
        Name & $\delta\beta$ & $\delta s$\\
        \hline
        $c_0$ & $1$ & $1$ \\
        $\epsilon_0$ & $1/\epsilon$ & $1$\\
        $\zeta_1$ & $\tau_e/\sqrt{\zeta}$ & $\tau_e$\\
        $\zeta_2$ & $1/\epsilon$ & $\sqrt{\zeta}/\epsilon$\\
        $\zeta_3$ & $1$ & $\sqrt{\zeta}$\\
        \hline
        \hline
        \end{tabular*}
    \end{center}
\end{table}

\newpage
\onecolumngrid

\begin{figure*}[t!]
    \centering
    \subfloat{
        \includegraphics[trim={0.0cm 0.0cm 0.0cm 0.0cm},clip, width=\textwidth]{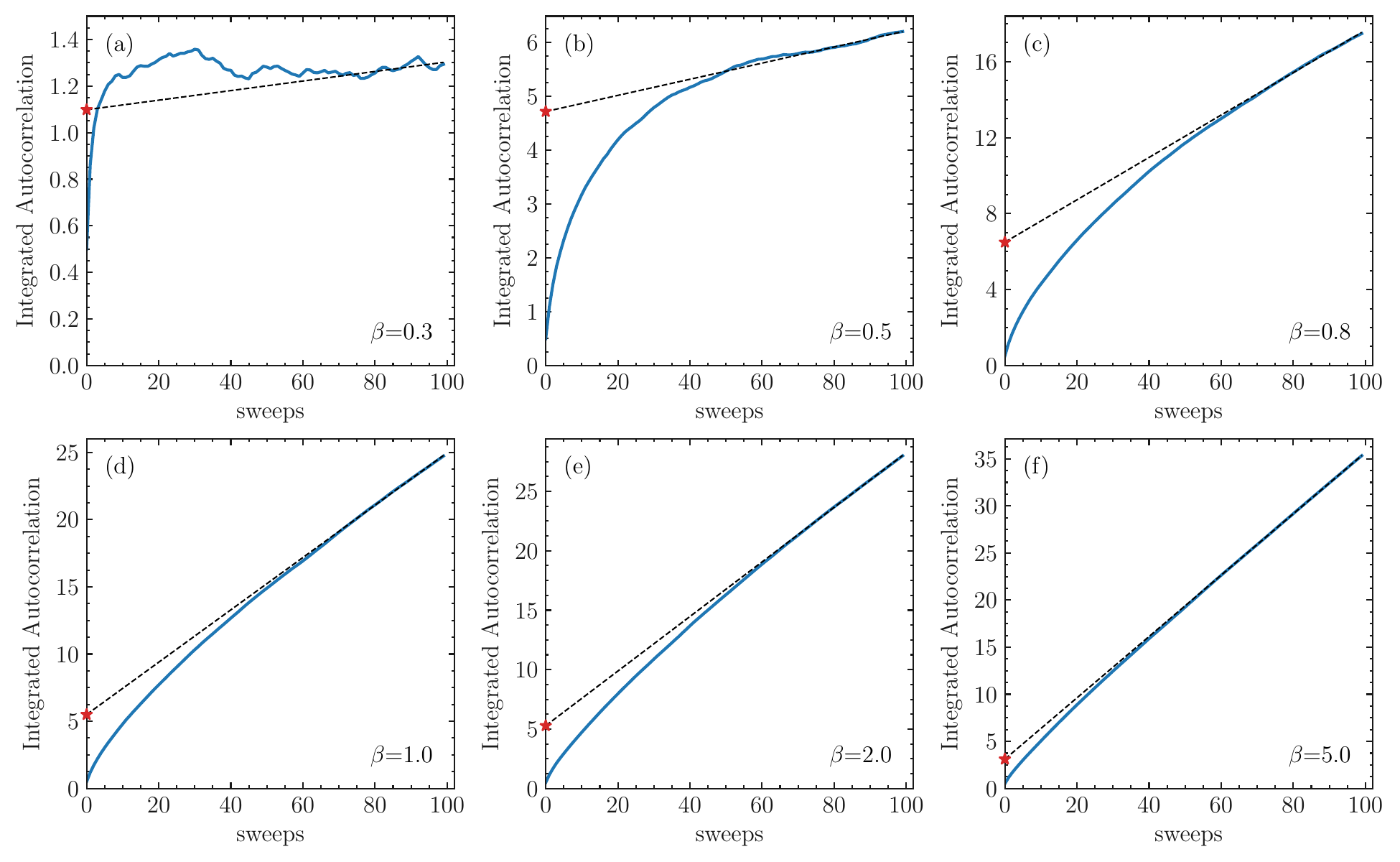}
    }
    \caption{Extraction of the energy autocorrelation time $\tau_e$ by linear fits to the integral of the energy autocorrelation function defined in Eq.~\eqref{eq:integrated-autocorrelation}. Panels (a)-(f) show examples of this procedure at different inverse temperatures $\beta$. We use a long run with 100 sweeps per temperature step for one realization in each bin. A linear curve is then fit to the data after 70 sweeps and the intercept (shown by the red stars) is interpreted as the autocorrelation time.}  
    \label{fig1:tau extraction}
\end{figure*}
\twocolumngrid

The five schedules and their designations are given in 
Table.~\ref{table1: schedule definitions}.  In all cases, the constants of proportionality are fixed by the total number of temperature steps (300) and the total number of sweeps, either 1000 or 3000.

\begin{figure}[t!]
    \centering
    \subfloat{
        \includegraphics[trim={0.4cm 0.1cm 0.0cm 0.1cm},clip, width=\columnwidth]{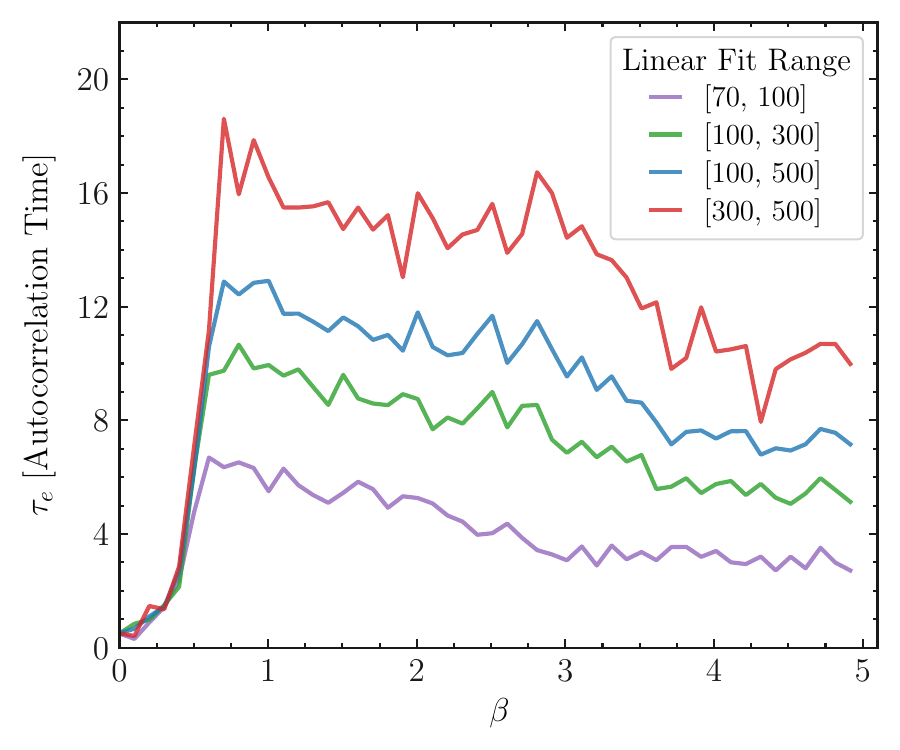}
    }
    \caption{Autocorrelation time $\tau_e$ versus inverse temperature $\beta$ extracted by fitting a line to different sections of the integrated autocorrelation function as illustrated in Fig.~\ref{fig1:tau extraction}. The integrated autocorrelation function was measured over 500 sweeps in total for a select disorder realization in bin 3. The noise in the curves is due to the fact that the data is taken from one single run here. 
    %We observe that different fit ranges yield significantly different results highlighting the fact the $\tau_e$ can only be measured up to a normalization factor. 
    For the performance comparison simulations, we used the bottom range [70, 100] (bottom curve) to define $\tau_e$.  }  
    \label{fig2: tau fit range}
\end{figure}

The 100 spin glasses samples used in our study were selected from a set of 5000 samples analyzed in Ref.\ \cite{AmMa18}.  As noted in \cite{AmMa18}, the distribution of hardness for PA, as measured by $\rho_f$ or $\rho_t$ is very broad and described by a log inverse Gaussian distribution.  The 100 samples used in our study were selected to span the whole range of hardness encountered in the randomly generated 5000 Gaussian instances.  Specifically, the 5000 instances were divided into 10 bins equally spaced on a logarithmic scale ranging from the easiest to the hardest instances. The 10 bins contain different numbers of instances but are of equal logarithmic width.  Ten instances were then randomly selected from each of the ten bins so that our study set of 100 instances are roughly uniformly spread over the range of (log) hardness that would be encountered in a much larger sample of instances.  The value of $\rho_t$ measured in \cite{AmMa18} is taken as the true measure of the hardness of each instance.  As observed in \cite{WaMaKa15b,AmMa18}, $\rho_f$ and $\rho_t$ differ by a relatively small constant for spin glasses and can be used interchangeably as a measure of computational hardness.  We note that resources devoted to the simulations in \cite{AmMa18} were much larger than used here and that we are reasonably confident that $\rho_t$ has been measured in the asymptotic regime and the exact ground state was found.

\subsubsection{Measuring the integrated autocorrelation time}

Relaxation times become very long in spin glasses so that we use the restricted friction tensor formalism developed for non-ergodic systems in Sec.\ \ref{sec:rzeta}, which, in any case reduces to the usual definition in the high temperature regime where equilibration is fast.  For simplicity of notation we refer to the restricted energy autocorrelation time, defined in Eq.\ \eqref{eq:rtau}, as $\tau_e$.
In order to obtain $\tau_e$  we measured the autocorrelation function in preliminary runs with  100 sweeps per temperature step for one disorder realization for each of the 10 bins. The sum of the energy autocorrelation function from 0 to  t sweeps, minus 1/2, was fit to a linear function of t in the interval 70 and 100 sweeps.  The intercept of the linear fit is identified as the restricted integrated energy autocorrelation time, $\tau_e$.  The value of $\tau_e$ used in the annealing schedule for each bin was obtained by averaging the value obtained from 20 runs on the examplar of that bin. 

Figure~\ref{fig1:tau extraction} shows the sum of autocorrelation function minus 1/2 for several values of $\beta$ for a disorder instance in bin 3 and a single long run.  The  red stars mark the location of the intercept, which is the value of $\tau_e$. For the highest temperature shown ($\beta=0.3$),  the integrated autocorrelation function is  noisy but appears to flatten indicating  that the full autocorrelation time has been measured.  However, for all larger values of $\beta$ the integrated autocorrelation function is nearly linear with a positive slope in the range 70 to 100 sweeps indicating that Metropolis dynamics are non-ergodic on these time scales but that a restricted integrated autocorrelation time can be measured.   

The lowest (purple) curve Fig.\ \ref{fig2: tau fit range}  shows $\tau_e$ as a function of $\beta$ from the same run as shown in Fig.\ \ref{fig1:tau extraction}.  We see that $\tau_e$ rises sharply as $\beta$ increases and reaches a maximum somewhat before the critical temperature.  Thereafter, in the non-ergodic low-temperature regime, $\tau_e$ slowly decreases.  The roughness of the curves is primarily because of the fact that this data is from a single run.

The simple assumption in Sec.\ \ref{sec:rzeta}  of two widely separated time scales does not hold for spin glasses where there are many time scales.  This fact manifests as an ambiguity in measuring $\tau_e$ depending on the range over which the linear fit is carried out.  Unfortunately $\tau_e$, which is the intercept of the linear fit, is very sensitive to the choice of the  fit window.
To demonstrate this point, we measured $\tau_e$ using different fit windows in the same long run as described above. The results are shown in Fig.~\ref{fig2: tau fit range}. 
As the window is moved toward longer times, there are two main effects: (1) the sharply rising curve for small $\beta$  extends closer to the critical point and reaches a higher value and (2) $\tau_e$ in the large $\beta$ phase is larger.   On the other hand, in the low temperature region,  the different curves for $\tau_e$ are similar except for a scale factor.  Thus, the primary effect moving the fitting window to larger times would be to sharpen the peak in the $\delta s / \delta \beta$ curve and to move computational work from the high temperature to the low temperature phase.  Given a limited total number of sweeps, it is not clear how to choose the best fitting window for measuring the restricted autocorrelation time. 

In addition to $\tau_e$, the variance of the energy, $\sigma_e^2$ is an input to the friction tensor. Our estimator of this variance is the geometric mean of population variance immediately after a resampling step and immediately before the next resampling step.  The temperature step size $\delta s$ for the fixed culling fraction schedule was obtained by direct measurement of $\delta s$ in the same simulations used to obtain $\tau_e$ and $\sigma_e^2$, which were carried out using the adaptive fixed culling fraction schedule described in Ref.\ \cite{AmMa18}.

\subsubsection{Annealing Schedules}

\begin{figure}[t]%[t!]
    \centering
    \subfloat{
        \includegraphics[trim={0.4cm 0.1cm 0.0cm 0.1cm},clip, width=\columnwidth]{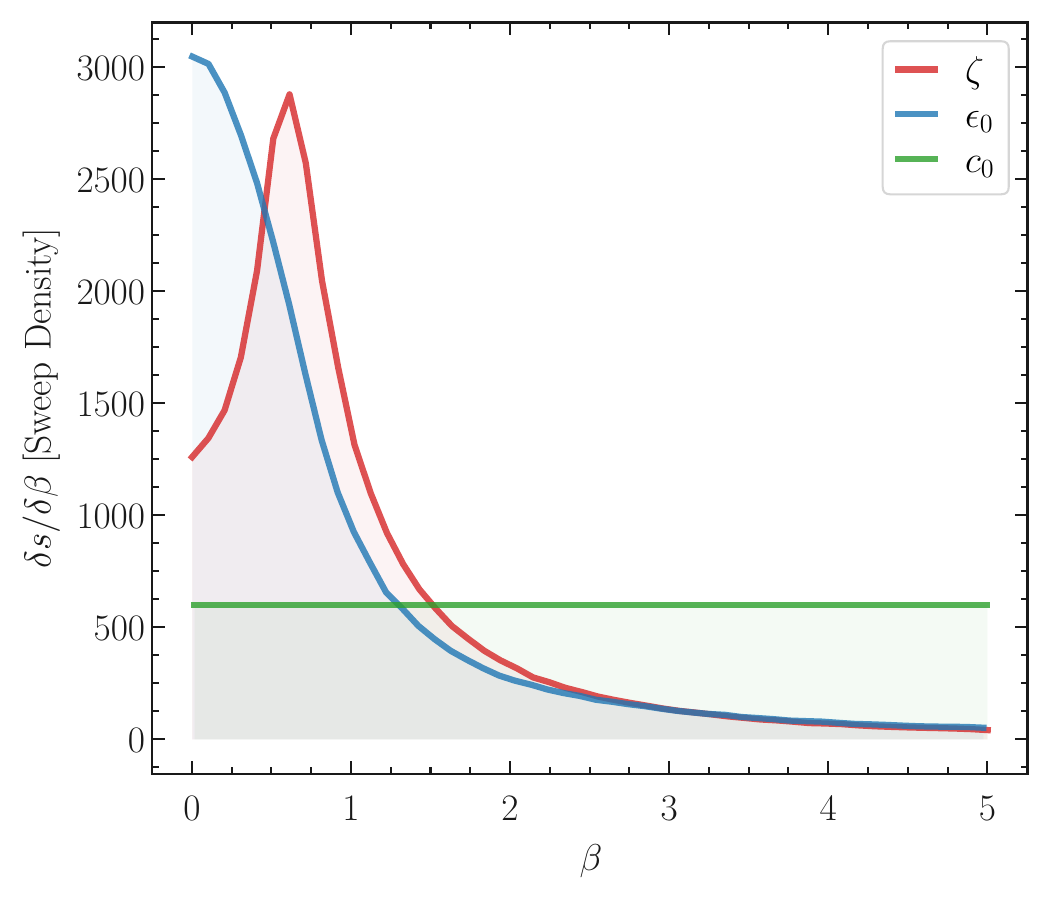}
    }
    \caption{Number of sweeps per unit $\beta$ as a function of inverse temperature $\beta$ for the three types of schedules: friction tensor (red, $\zeta$), fixed culling fraction (blue, $\epsilon_0$), and constant $\beta$-step (green, $c_0)$.  The total number of sweeps, i.e., the area under the curves is $3000$ in all three cases. The $c_0$ schedule distributes the sweeps uniformly in temperature, whereas the $\epsilon_0$ and $\zeta$ schedules put more work at high temperatures and close to the critical point, respectively. }  
    \label{fig3:schedules}
\end{figure}

Figure~\ref{fig3:schedules} shows $\delta s/\delta \beta $ for the three types of schedule. A single curve labeled $\zeta$ can be used for all three friction tensor schedules since they all have the same functional form for $\Delta s/\Delta \beta$.  The constant culling fraction schedule and the friction tensor schedules are similar in doing very little computational work at large $\beta$ and instead relying on resampling.  The friction tensor schedule differs from the fixed culling schedule in doing more sweeps in the region near the spin glass phase transition (at $\beta=1.07$, see Ref.\ \cite{KaKoYo06}), where the autocorrelation time becomes large.  The idea that more sweeps are required near a phase transition or where the autocorrelation time becomes large has been discussed in other studies of annealing and tempering algorithms \cite{AnGo94,KaTrHuTr06,BiNuJa08,AmMa18,BaPaWaKa18}. Figure~\ref{fig3:schedules} shows the fixed culling and friction tensor schedules for bin 3 but in practice the schedules for all bins were roughly the same and it would have been sufficient to choose a single schedule for all bins.  

The culling fraction is roughly proportional to the standard deviation of the energy (see Eq.\ \eqref{eq:eps12gaussian}), which explains why it decreases monotonically with $\beta$.  The friction tensor is proportional to the product of the variance of the energy and the restricted integrated autocorrelation time, which leads to a sharply peaked functional form for $\delta s/ \delta \beta$ near the critical point. 

\subsection{Performance of Annealing Schedules}
\label{sec:results}
\begin{figure*}[t!]
    \begin{center}
        \subfloat{
            \includegraphics[trim={0.1cm 0.1cm 0.0cm 0.1cm}, clip, width=0.9\textwidth]{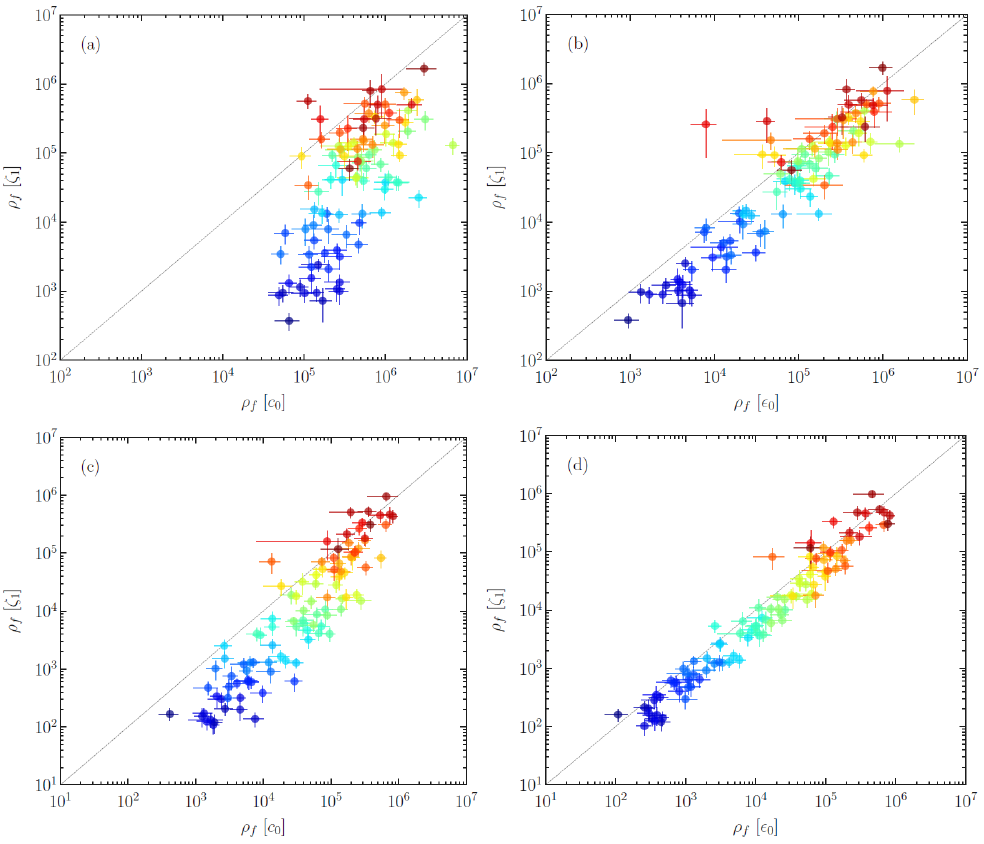}
        }
        \hfill
        \subfloat{
            \includegraphics[trim={0.0cm 0.2cm 0.0cm 0.2cm}, clip, width=0.6\textwidth]{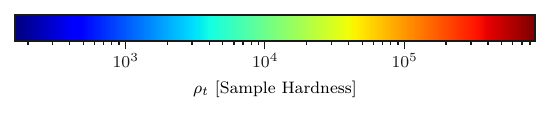}
        }
        \caption{Scatter plot comparison of $\rho_f$ for the friction tensor schedules $\zeta_1$ versus the flat schedule $c_0$ and constant culling schedule $\epsilon_0$. Each dot shows one sample with the vertical position of the dot representing $\rho_f$ for the $\zeta_1$ friction tensor schedule and the horizontal position representing the competing schedule, either constant $\beta$-step, $c_0$ or constant culling fraction $\epsilon$. The color of the dot is the value of $\rho_t$, given in the color bar below the panels, for that sample obtained from long simulations \cite{AmMa18} and represents the hardness of the corresponding instance. Easy samples are in the lower left, while the hard samples are concentrated in the upper right of each panel. Panels (a) and (b) are measured in  1000-sweep simulations, whereas (c) and (d) in 3000-sweep simulations.  The value of $\rho_f$ is not meaningful when it is greater than $10^5$ so the comparison should only be trusted for the easier samples. The friction tensor schedule outperforms the other two schedules although the difference is smaller as more sweeps are used. }
        \label{fig4:schedule rho_f comparison}
    \end{center}
    \vspace{-0.3cm}
\end{figure*}

Figure.~\ref{fig4:schedule rho_f comparison} presents $\rho_f$ values for the 100 disorder realizations comparing the three types of schedules.  Each panel is a scatter plot comparing a friction tensor schedule to one of the two other types of schedules and the coordinates of each dot represent the values of $\rho_f$ for a single disorder realization and the two schedules labeled on the axes.  The upper panels are results from 1000 sweep simulations and the lower panels form 3000 sweep simulations. The easiest samples are found in the lower left and the hardest in the upper right of each panel. We have shown only the $\zeta_1$ friction tensor schedule.  Similar scatter plots comparing the three friction tensor schedules are not shown because the $\rho_f$ values are statistically indistinguishable.

When $\rho_f$ is comparable to or larger than $R$ the system is not well-equilibrated and the value of $\rho_f$ is likely to be underestimated, both because the population size is too small and the number of independent runs to measure it is too small.  For these reasons, $\rho_f$ is best suited for comparing the performance of different schedules for the easier samples.   We see in panel (a) that $\rho_f$ for the constant $\beta$-step schedule is always near or above $10^5$  suggesting it is unable to well-equilibrate any of the samples with 1000 sweeps. On the other hand the constant culling and friction tensor schedules do succeed in equilibrating most of the easier samples with 1000 sweeps.  For the hardest samples, none of the schedules reach equilibrium even with 3000 sweeps. For the harder samples, we cannot draw meaningful comparisons between schedules based on the $\rho_f$ measurements.

It is clear from Fig.~\ref{fig4:schedule rho_f comparison} that the friction tensor and fixed culling schedules are far superior to the fixed $\beta$ schedule for the easier samples.  From panels (b) and (d) we see that the friction tensor schedule outperforms the fixed culling schedule though by a relatively small margin.  Finally, it is interesting to observe that the differences between the schedules are most pronounced for the 1000 sweep simulations where computational resources are more limited.

We next turn to the probability of finding the ground state.  This metric is better suited to comparing the performance of the schedules for the harder samples in the study set.  These samples are typically not well-equilibrated at the lowest temperature, nonetheless, the ground state is sometimes found.

\begin{figure*}[h!]
    \centering
    \subfloat{
        \includegraphics[trim={0cm 0cm 0cm 0cm},clip, width=0.97\textwidth]{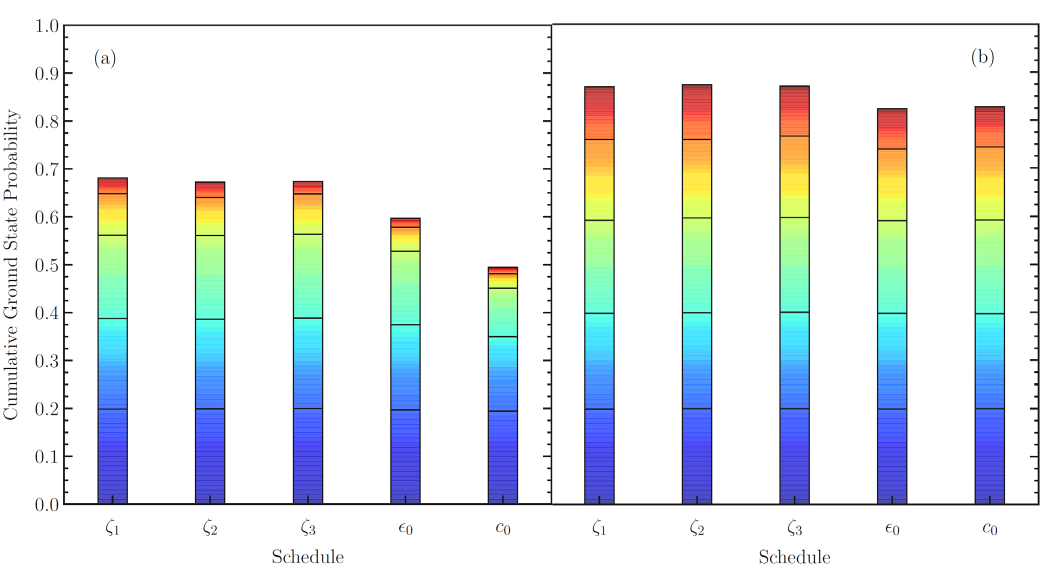}
    }
    \hfill
        \subfloat{
            \includegraphics[width=0.6\textwidth]{figures/continuous_horizontal_colorbar.pdf}
        }
    \caption{Cumulative probability of finding the ground state for the five annealing schedules. The height of each bar represents the overall probability of finding the ground state for the 100 disorder samples.  Each bar is divided by horizontal lines into 5 blocks each containing 20 disorder realizations whose hardness is represented by the color bar with the easiest problems in the bottom block and hardest problems in the top block. The left panel (a) shows results from 1000 sweep simulations and the right panel (b) 3000 sweep simulations.  The ground state is almost always found using all schedules for the easier problems but for the hardest problems the ground state is rarely found. }
    \label{fig5:schedule success}
\end{figure*}

Figure.~\ref{fig5:schedule success} is a bar graph showing the cumulative probability of finding ground states for all 100 samples for each of the three types of schedules.  If an algorithm always found the ground state for every sample, the height of the bar would be one.  The easiest samples are at the bottom of the bar and the height of the bar at a given color level is the cumulative probability of finding ground states easier than the hardness represented by that color.  The horizontal lines in each bar separate groups of two hardness bins.  Panel (a) presents results for 1000 sweeps and panel (b) for 3000 sweeps.

First note the ground states are found with very high probability by all schedules for the two easiest bins (i.e. the height of the first horizontal line is near 0.2, the value expected if the ground state is found for all 20 samples in all 20 runs for that sample).  On the other hand, for the hardest 20 samples, represented by the small rectangles at the top of each bar, the probability of finding ground states is much less than 0.2.  As was the case in comparing the schedules by $\rho_f$  the friction tensor schedule performs best and the constant $\beta$-step schedule worst.  All algorithms improve in going from 1000 to 3000 sweeps and the difference between the algorithms diminishes.  The most dramatic differences are found for the probability of finding the ground state for the 20 hardest samples using 1000 sweeps.  The average probability per sample of finding the ground state is 0.16 for the friction tensor schedule, 0.09 for the fixed culling schedule, and 0.06 for the constant $\beta$-step schedule.

%\newpage
\onecolumngrid
%\twocolumngrid
\newpage
%\onecolumngrid

%\clearpage %\newpage
\twocolumngrid
\section{Summary and Discussion}
\label{sec:conclude}

We have developed a formalism for quantifying the performance of annealing schedules that is applicable to population annealing and annealed importance sampling.  This formalism applies to annealing schedules that are sufficiently slow that the system remains near equilibrium throughout the process.  In this setting the performance metric for annealing algorithms, $\rho_f$ is formally equivalent to the friction tensor metric introduced by Sivak and Crooks \cite{SiCr12} for optimal control of thermodynamic systems.   In the usual case of annealing in a single parameter (inverse temperature) an explicit equation determines the rate of change of the parameter during the anneal.  The result is that the velocity (typically Monte Carlo sweeps per unit time) is proportional to the inverse of the square root of the friction tensor.  The friction tensor itself is the product of the variance of the energy and the energy autocorrelation time.  

In the interesting case of annealing in a multi-dimensional parameter space the optimal annealing path is a geodesic in the metric defined by the friction tensor  \cite{SiCr12}.  Identifying geodesics or approximate geodesics in realistic multi-parameter annealing situations is a topic for future work. Multidimensional annealing schedules could play an important role in constrained optimization where annealing parameters can be used to control the penalties associated with constraints.

In addition to near-equilibrium annealing, we also considered the common situation of ergodicity breaking associated with first-order transitions or temperature chaos.  In these situations, the friction tensor formalism must be supplemented by a term associated with traversing the ergodicity-breaking region. In a first-order transition, there are two (or more) local minima in the free energy landscape and the transition occurs where the global minimum changes from a single "high temperature" local minimum to one (or more) "low temperature" minima.   At some point in parameter space before the transition, ergodicity breaking generally occurs.  In the context of a simple two-state model, we calculated that contribution to $\rho_f$ is equal to the inverse of the probability of being in the low-temperature minima when ergodicity breaking occurs.  In addition, we proposed a modified version of the correlation time matrix to include the presence of broken ergodicity.  Overall, optimizing a multi-dimensional annealing path involves minimizing the sum of the friction tensor contributions from the continuous path segments and the contribution of the ergodicity-breaking transitions.  It is left for future work to study how to deploy these ideas to find good annealing paths for a realistic multi-parameter optimization problem.

Although the formalism applies to smooth annealing schedules that remain near equilibrium except at sharp ergodicity-breaking transitions, we expect it to be useful in more realistic settings.  To explore the practical validity of these ideas,
we carried out numerical simulations on the three-dimensional Ising spin glass using population annealing with inverse temperature as the sole annealing parameter.  We found that an annealing schedule based on the friction tensor (modified for ergodicity breaking) outperformed schedules based on either a fixed culling fraction or constant steps in inverse temperature. The constant $\beta$-step schedule performed very poorly related to the other two schedules because it allocates too much work to the Metropolis algorithm at low temperatures where both the variance of the energy and the modified autocorrelation time are small.  On the other hand, the fixed culling fraction schedule, which has been employed in previous population annealing studies, performs almost as well as the friction tensor schedules. These two schedules turn out to be closely related.  The difference is that the friction tensor includes a factor of the energy autocorrelation time and the culling fraction does not.  The good performance of the fixed culling fraction schedule results perhaps from the small variation of the energy autocorrelation time when modified for ergodicity breaking.  Given the simplicity of determining the culling fraction on the fly, a fixed culling fraction schedule is useful in practice. It may be that a fixed culling fraction schedule will also prove useful in the context of multi-parameter annealing.

\acknowledgements{We thank Arnaud Doucet, Helmut Katzgraber, Brad Lackey and Martin Weigel for useful discussions.}

\appendix
\section{Pairwise Residual Resampling}
\label{app:limit}

To carry out pairwise residual sampling (\prr),  the set of replicas is divided into two subsets, $S^+$ and $S^-$ according to whether the replica weight is greater or less than one,  $S^+ = \{r | \wt_r>1 \}$  and $S^- = \{r | \wt_r < 1 \}$. (It may be the case that some replicas have $\wt=1$.  These replicas are untouched by \prr.)  We define the total positive and negative \textit{residual} weights $W^+$and $W^-$ associated with $S^+$ and $S^-$, respectively, as,
\begin{equation}
    W^{\pm}= \sum_{r \in S^{\pm} } |\wt_r -1|,
\end{equation}
We make the assumption that the total residual weight is less than one $(W^++W^-)<1$, which will hold when $dt$ is sufficiently small. Individual residual weights are then also less than one and since the sum of all $\wt$'s is $R$ we must have that
$W^+ = W^- \equiv W$.  Note that the expected value of $W/R$ is the culling fraction (see Eq.\ \eqref{eq:eps}).

Because the positive and negative residual weights are equal, we are allowed to resample in pairs, making an extra copy of a replica in $S^+$ and culling a replica in $S^-$ so that the population size stays fixed.  Furthermore, in the limit $W < 1 \ll R$, no changes to the population are made with probability $1-W/R$ or one pair is copied/culled with probability $W/R$.  Resampling involving more than one pair is suppressed by an extra factor of $W/R$ and can be ignored in the large $R$ limit.  In the case that the culling/copying event is chosen, the culled replica, $r_-$ is chosen from $S^-$ with probability $(1-\wt_{r_-})/W$ and the copied replica $r_+$ chosen from $S^+$ with probability  $(\wt_{r_+}-1)/W$.  
 It is straightforward to verify that requirement for the copy numbers, Eq.\ \eqref{eq:ntau}, is satisfied and that, like nearest integer resampling, for every replica in $S^+$,  $n_r$ is either 1 or 2 while for every replica in $S^-$, $n_r$ is either 0 or 1.

 We now show that PRR is well-defined for arbitrarily small $dt$ and that the overall contribution of resampling to the variance of observables is negligible.  Consider the variance of the population-averaged congugate forces,  $\Xt_k$. These variances, through the Schwarz inequality, bound the contributions to the growth of $\rho_f$. Our aim is to show that the contribution to the growth of these variances due to resampling is negligible in the small $dt$ and large $R$ limit.  
 
 In a single resampling step, $\Xt_k$ is modified according to, 
\begin{equation}
\label{eq:prrxt1}
    \Xt_k(t^+) = \frac{1}{R}\sum_r n_r X_k^r(t),
\end{equation}
which can be re-written as,
\begin{equation}
\label{eq:prrxt}
    \Xt_k(t^+) = \Xt_k(t) + \frac{1}{R}\sum_r (n_r-1) X_k^r(t).
\end{equation} 
These equations can now be used to analyze the evolution of $\var(\Xt_k(t))$ in the absence of MCMC dynamics.

Using the Law of Total Variance we have that 
\begin{align}
\label{eq:prrvar}
\var(\Xt_k(t^+)) &= \var(\E (\Xt_k(t^+)|\{\Xb^r(t)\} )) \\ \nonumber
&+\langle \var(\Xt_k(t^+) | \{\Xb^r(t)\})\rangle_\Lambda , 
\end{align}
where the conditional expectation and conditional variance are carried out holding the population at time $t$ fixed but allowing randomness in the choice of copy numbers, $\{n_r\}$ given the weights $\{\wt_r\}$. The unconditioned variances are with respect to the full non-equilibrium distribution induced by PA.  Using Eq.\ \eqref{eq:prrxt1} and the fact that the expection of $n_r$ is $\wt_r$, the first term on the RHS of Eq.\ \eqref{eq:prrvar}  can be written as
\begin{equation}
   \var(\E (\Xt_k(t^+)|\{\Xb^r(t)\} ))= \frac{1}{R^2} \var(\sum_r \wt_r X_k^r),
\end{equation}
where in this equation and henceforth in the appendix, a variable without a time argument will assumed to be at the time immediately before the resampling event $t$. 

In the small $dt$ limit we can use Eq.\ \eqref{eq:taudt} and expand to leading order in $dt$ to obtain
\begin{equation}
     \var(\E (\Xt_k(t^+)|\{\Xb^r\} ))= \frac{1}{R^2} \var(\sum_r (1-dt \dot{ \lambda}_i (X_i^r - \tilde{X}_i)) X_k^r )
\end{equation}
If the population is close to equilibrium and uncorrelated we can obtain the approximate expression,
\begin{align}
    \var(\E (\Xt_k(t^+)|\{\Xb^r(t)\} )) & = 
    \var(\Xt_k) \\ \nonumber & + \frac{2}{R} dt \dot{\lambda}_i \Cov( \delta X_i X_k  , X_k ),
\end{align}
where the covariance is respect to the Gibbs distribution at parameter values $\lamb(t)$.  We will show below that the second term in the decomposed variance Eq.\ \eqref{eq:prrvar} is of negligible magnitude compared to $\var(\Xt_k)$. This enables us to re-write the above relation as the differential equation
\begin{equation}
    \frac{d \var{(\Xt_k)}}{dt} = \frac{2}{R} \dot{\lambda}_i \Cov( \delta X_i X_k  , X_k ),
\end{equation}
which describes the change in variances due to the change in the parameters of the Gibbs distribution.  This equation is not exact because of correlation in the population but it reveals the order in $R$ dependence.  The  $1/R$ behavior of $\var(\Xt_k(t))$ is expected since $\Xt_k$ is an average over a population of size $R$.

The second term on the RHS of Eq.\ \eqref{eq:prrvar} contains the contribution to  $\var(\Xt_k)$ arising from resampling.  Our goal is to show that this term can be integrated in the $dt \rightarrow 0$ limit and is higher order in $1/R$.  From Eq.\ \eqref{eq:prrxt} and noting that $\Xt_k(t)$ is fixed by the conditioning we have 
\begin{equation}
  \langle \var(\Xt_k(t^+) | \{\Xb^r\})\rangle_\Lambda = \frac{1}{R^2}\langle \var(\sum_r (n_r-1)X_k^r | \{\Xb^r\})\rangle_\Lambda  
\end{equation}
The argument of the variance on the RHS of this equation is a random variable that is equal to zero with probability $1-W/R$ and equal to $(X_k^{r_+}-X_k^{r_-})$ with probability $W/R$.  Note that the argument of the variance involves at most two terms in the sum over the population.  The culling probability (see Eq.\ \eqref{eq:eps}) is an infinitesimal given by
\begin{equation}
    W/R = \frac{dt}{2R}    \sum_r |\dot{ \lambda}_i(X^r_i -\Xt_i)|.
\end{equation}
and near equilibrium can be approximated by $dt \langle |\dot{ \lambda}_i\delta X_i| \rangle_{\lambda(t)} $. 

These observation lead to the conclusion that the resampling contribution to the increase in variance is order $dt$ so it can be integrated and is well-defined in the $dt \rightarrow 0$ limit.
Furthermore, the contribution from resampling is order $1/R^2$ whereas $\var(\Xt_k)$ is order $1/R$ so the resampling contribution is negligible in the large $R$ limit.

\bibliographystyle{unsrt}
\bibliography{references}
\end{document}